\documentclass{moriond}

\def\be{\begin{equation}}
\def\ee{\end{equation}}
\def\bea{\begin{eqnarray}}
\def\eea{\end{eqnarray}}

\newcommand{\Photo}{}

\begin{document}
\vspace*{4cm}
\title{$B\to D^\ast\ell\nu$ from LQCD: is there light at the end of the tunnel?}

\author{Alejandro Vaquero \\ On behalf of the Fermilab Lattice and MILC collaborations}

\address{Departmento de Física Teórica, Universidad de Zaragoza,\\
C/ Pedro Cerbuna 12, 50009 Zaragoza, Spain. \\
Centro de Astropartículas y Física de Altas Energías,\\
C/ Pedro Cerbuna 12, 50009 Zaragoza, Spain}

\maketitle\abstracts{
Lattice QCD (LQCD) calculations play a key role in the establishment of flavor anomalies.
One of the most recent advancements in LQCD to this end has been the publication of several calculations of the $B\to D^\ast\ell\nu$ form factors,
but despite all the anticipation, the LQCD results have been unable to give a final answer to the questions it was destined to answer.
In this work I briefly review what is the current status of heavy-to-heavy and heavy-to-light semileptonic decays calculations in LQCD, and what we can expect for the near and not-so-near future.}

\section{Introduction and motivation}
The Intensity Frontier has become a cornerstone of beyond the Standard Model (BSM) physics searches.
The reason is the high sensitivity of the indirect searches to new phenomena at large scales, as opposed to the direct searches that require larger and more powerful machines to explore.
The flavor sector of the Standard Model (SM) is rich in opportunities to find BSM physics.
In particular, the $B$~anomalies---understanding here anomaly as a tension between SM calculations and experimental measurements---are an important source of candidates where we expect to find new physics.
There are two outstanding candidates within the $B$~anomalies: the tensions in the CKM matrix elements, and the lepton-flavour universality (LFU) ratios.

\begin{figure}[h!]
\centering                  
\includegraphics[width=0.39\textwidth,angle=0]{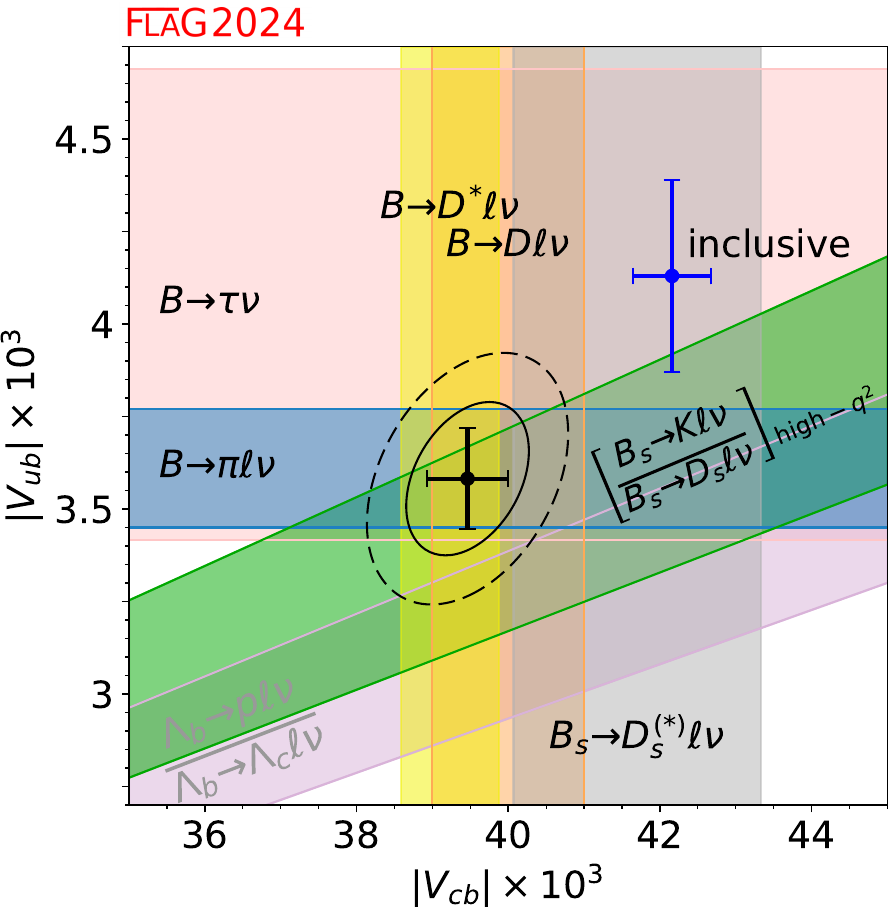}
\includegraphics[width=0.59\textwidth,angle=0]{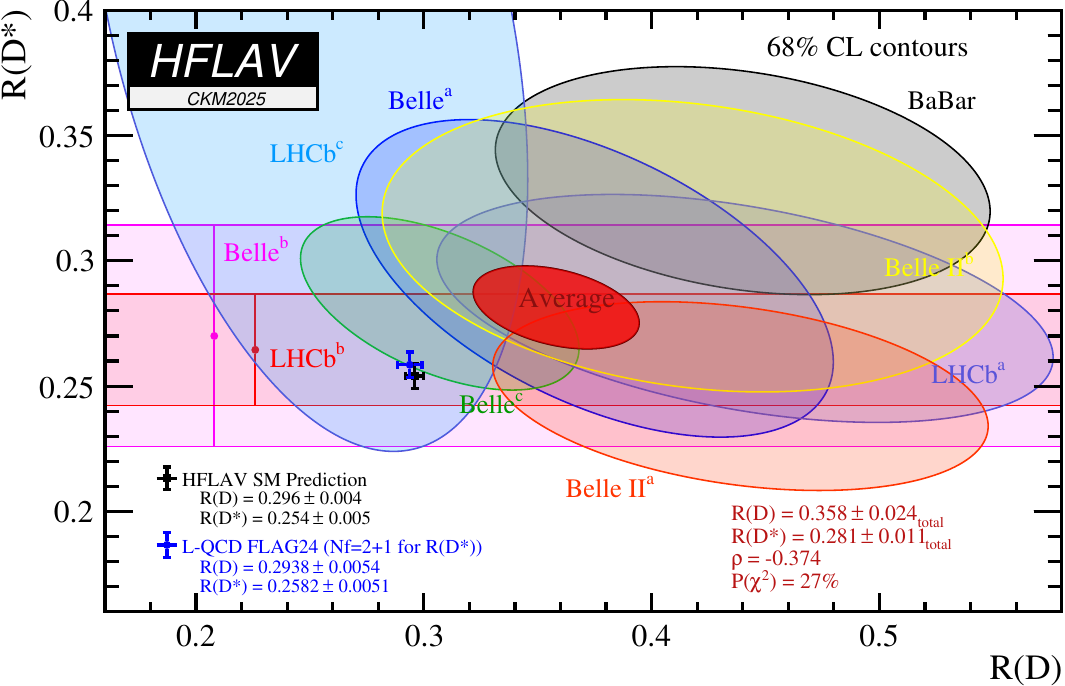}
\caption{Tensions in $|V_{xb}|$ (left) and in $R(D^{(\ast)})$ (right), according to FLAG 2024 and HFLAV.}
\label{FLAG-Vxb}            
\end{figure}

\begin{figure}[h!]        
\centering                  
\includegraphics[width=\textwidth,angle=0]{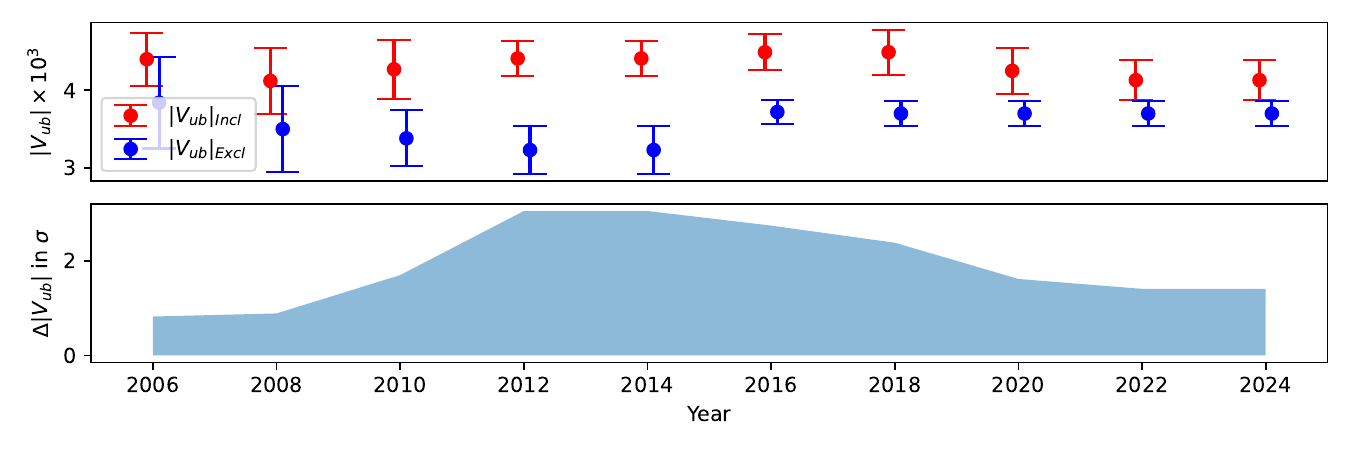}
\caption{{\bf Upper panel:} Evolution of the tension between inclusive and exclusive $|V_{ub}|$ as a function of the year, from the PDG reports.
         {\bf Lower panel:} Difference in $\sigma$ between the inclusive and the exclusive results.}
\label{VubDiff}
\end{figure}

\begin{figure}[h!]        
\centering                  
\includegraphics[width=\textwidth,angle=0]{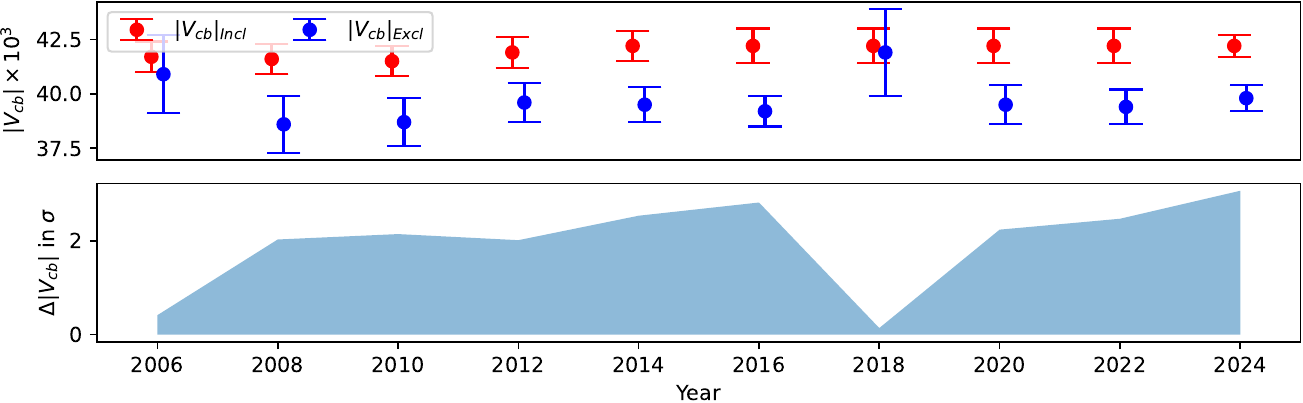}
\caption{{\bf Upper panel:} Evolution of the tension between inclusive and exclusive $|V_{cb}|$ as a function of the year, from the PDG reports.                
         {\bf Lower panel:} Difference in $\sigma$ between the inclusive and the exclusive results.}
\label{VcbDiff}
\end{figure}

There is a long-standing tension between the inclusive and the exclusive determinations of the $V_{xb}$ CKM matrix elements.
The origin of this tension is not completely understood, but it has been the subject of discussion for decades.
The current status of the discrepancies, according to FLAG~\cite{FlavourLatticeAveragingGroupFLAG:2024oxs}, is shown in the left pane of fig.~\ref{FLAG-Vxb}.
The tension shown is not equally split between $V_{ub}$ and $V_{cb}$.
Figure~\ref{VubDiff} shows the evolution of the inclusive-exclusive tensions in $V_{ub}$ over the years, according to the PDG~\cite{ParticleDataGroup:2006fqo,ParticleDataGroup:2008zun,ParticleDataGroup:2010dbb,ParticleDataGroup:2012pjm,ParticleDataGroup:2014cgo,ParticleDataGroup:2016lqr,ParticleDataGroup:2018ovx,ParticleDataGroup:2020ssz,ParticleDataGroup:2022pth,ParticleDataGroup:2024cfk}, and shows that tne matrix element $V_{ub}$ is under control: the inclusive-exclusive difference has been decreasing lately, and the current tension is under $2\sigma$.
The story for $V_{cb}$, shown in fig.~\ref{VcbDiff}, looks quite different.
There has been a mild tension of about $2\sigma$ or $3\sigma$ since 2008, and we have not yet understood why\footnote{But at least we have understood the dip in 2018~\cite{Bigi:2017njr,Grinstein:2017nlq,Waheed:2018djm,Dey:2019bgc}.}.
The latest developments in lattice QCD (LQCD) and experiment (see for instance~\cite{FermilabLattice:2021cdg,Aoki:2023qpa,Harrison:2023dzh,Belle-II:2025rna,LHCb:2024jll}) have instilled hope in solving this tension, but the progress has been slower than desired.
On the other hand, the last years have also brought improvements in the inclusive calculation~\cite{Bordone:2021oof,Bernlochner:2022ucr,Finauri:2023kte}.
None of these advancements have not been enough to elucidate the question of $V_{cb}$.
                            
The other interesting place to look for new physics is the LFU ratios, whose current status, according to HFLAV~\cite{HFLAV:2024ctg,Belle-II:2024ami}, is shown in the right pane of fig.~\ref{FLAG-Vxb}.
The tensions between theory and experiment are at the level of $\approx 3-4\sigma$, and come mainly from the $R(D)$ contribution.
Even though this is a sizable tension, it is still under the discovery threshold.
Thus, it is urgent to improve our existing results to find out the fate of these $B$~anomalies.

\section{A short introduction to lattice gauge field theories}
Lattice gauge field theories are based on the Feynman path integral formalism, which can be understood as a weighted sum of possible trajectories.
The weights are complex phases, but a Wick rotation allows us to convert these phases into real exponentials, and thus interpret the weights as unnormalized probabilities in a partition function.
A lattice calculation consist of an ensemble of randomly generated configurations of the gauge and fermion fields, following the appropriate probability distribution function --which is given by the QCD action--, and the relevant correlation functions computed on each configuration.
Then one can compute averages and extract physical observables.

The ensembles and correlation functions are generated in supercomputers.
To this end, the QCD action is regularized by discretizing spacetime, and then the system is put in a finite box.
The UV regulator is the lattice spacing $a$, and one can recover physical results by pushing the regulator to infinity, i.e., by taking the limit $a\to 0$, which is also called the continuum limit.
The finite box is necessary, so the number of degrees of freedom to simulate is finite, and they fit in the computer.
Finite volume effects are usually negligible in most applications, including heavy-to-heavy decays, but some calculations require a careful analysis of these effects.

The main problem in heavy meson decays is the fact that the bottom quark mass usually lies beyond the UV regulator.
This introduces large discretization errors coming from the heavy quark sector, which are translated into large systematic errors.
In fact, this is often the largest contribution to the systematic errors in the LQCD calculations of heavy decays.
There are two approaches to deal with this problem: the first one uses an effective field theory (EFT) to simulate the heavy quarks.
We can then easily reach physical heavy quark masses, but the matching between LQCD and the EFT introduces sizable systematic errors.
The second approach is to use an improved fermionic action, that reduces discretization errors, and reach very fine lattice spacings.
Even in that case, $m_b$ is normally out of reach, and we need to perform simulations at unphysical values of the bottom quark mass, and then extrapolate to the right value.

\section{The current mess in LQCD calculations}
\subsection{Heavy-to-heavy decays}
\begin{figure}[h!]        
\centering                  
\includegraphics[width=0.455\textwidth,angle=0]{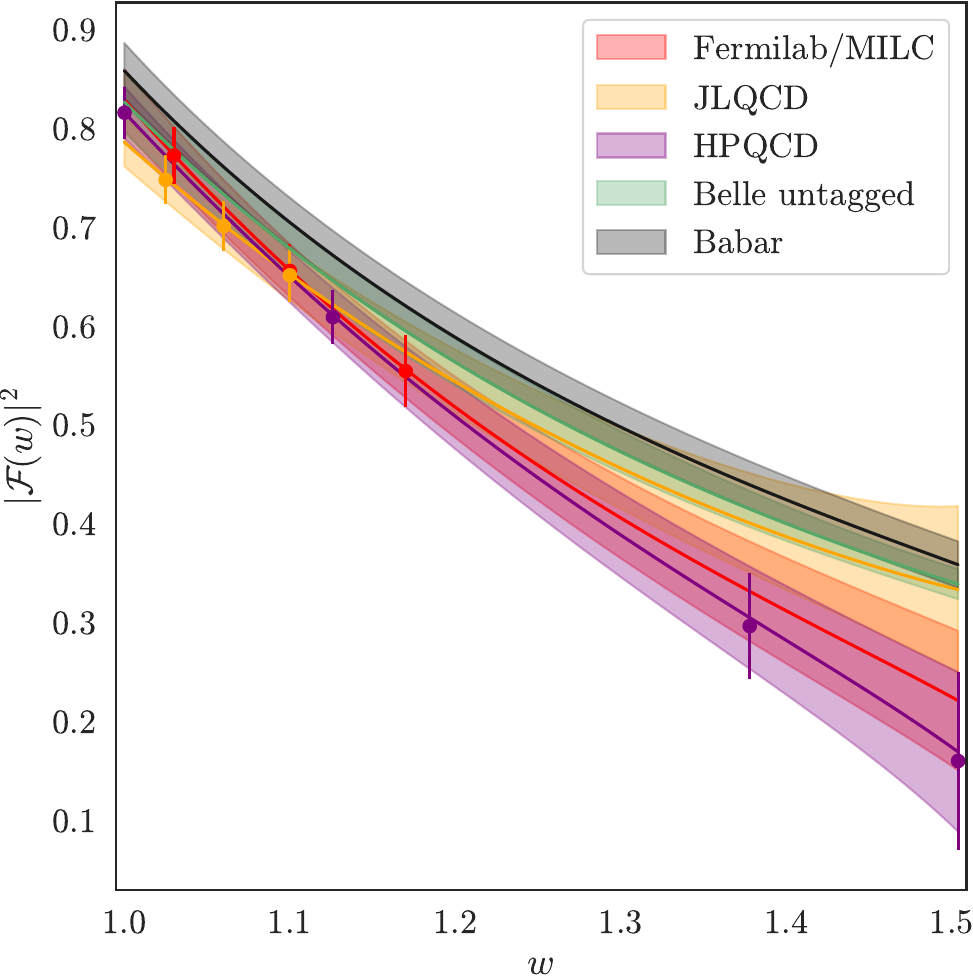}
\caption{Comparison of the shape of the decay amplitude $\eta_{\textrm{EW}}\left|V_{cb}\right|^2\left|\mathcal{F}(w)\right|^2$ among different LQCD calculations, and Belle and BaBar experimental measurements.}
\label{LQCD-BtoDst}         
\end{figure}

Arguably, the most relevant development in LQCD calculations of heavy-to-heavy deacys has been the publication of three, completely independent, lattice-QCD calculations of the $B\to D^\ast\ell\nu$ form factors~\cite{FermilabLattice:2021cdg,Aoki:2023qpa,Harrison:2023dzh},
featuring different fermion actions, and hence different systematic errors.
Figure~\ref{LQCD-BtoDst} shows the current results for the decay amplitude of the decay, including a comparisong with Belle~\cite{Waheed:2018djm} and BaBar~\cite{Dey:2019bgc} results.
Even though the three LQCD results agree well within $2\sigma$, some calculations display a mild tension with experiment.
These small differences have sparked diverse speculations and discussions in workshops and conferences.
Certainly, the LQCD results might look messy, but the agreement among them is objectively good.
One can perform combined fits of the data to obtain more precise results than those offered by a single collaboration, and the quality of the resulting fit is high, no matter which combination of LQCD results is selected.
The same cannot be say about experimental data: a combined BGL fit of the BaBar and Belle data sets for the three form factors involved in this decay amplitude results in a clear tension, with a $p$~value of order $10^{-2}$.
One can understand the dispersion between the different LQCD results as a systematic error, and although the situation is under control, the LQCD community should be careful in their analysis of their systematic errors for the next calculations,
to prevent this healthy dispersion from becoming a real issue.
Hence, it is clear that we need more and better calculations of the form factors of this channel.

\subsection{Heavy-to-light decays}
\begin{figure}[h!]        
\centering                  
\includegraphics[width=0.49\textwidth,angle=0]{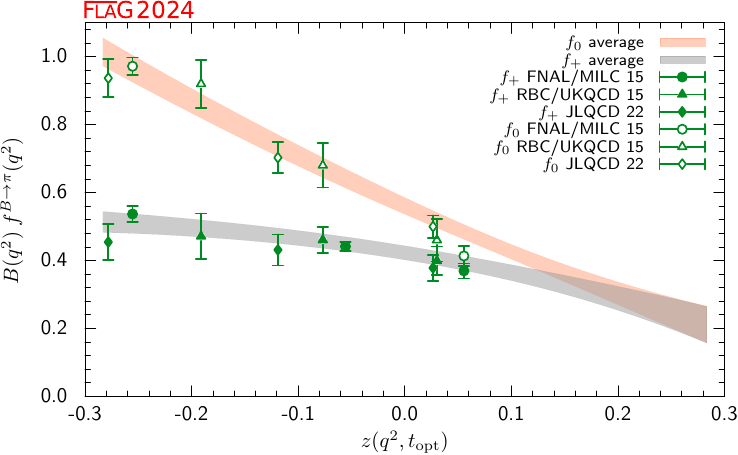}
\includegraphics[width=0.49\textwidth,angle=0]{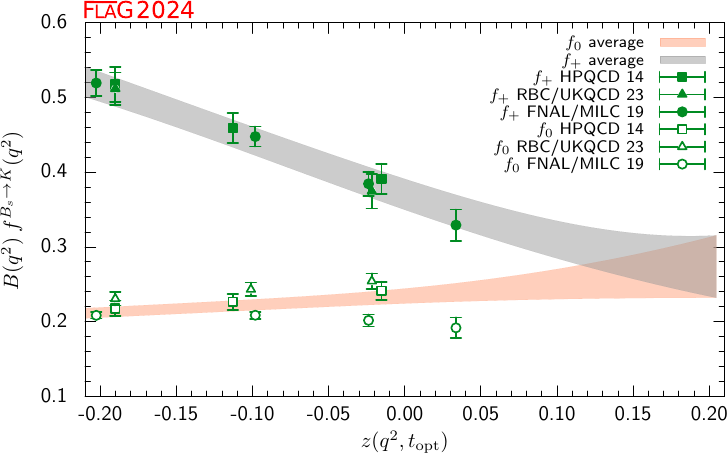}
\caption{Tensions in the $B\to\pi\ell\nu$ (left) and the $B_s\to K\ell\nu$ form factors, according to FLAG 2024.}
\label{LQCD-BtoLight}       
\end{figure}

Whereas in the $B\to D^\ast\ell\nu$ channel the form factors show a healthy dispersion, well within compatibility bounds, the heavy-to-light channels show real tensions among different LQCD form factors that should be addressed.
The latest FLAG report highlight serious circumstances with the $B\to\pi\ell\nu$ and $B_s\to K\ell\nu$ form factors~\cite{FlavourLatticeAveragingGroupFLAG:2024oxs}, as shown in fig.~\ref{LQCD-BtoLight}.
Since FLAG is the resource experimentalist and phenomenologist use to access lattice-QCD results, the inconsistency might result in a lower confidence in the data generated by the lattice community.
It is of foremost importance to find an explanation to this issue, as well as to supersede existing calculations by better ones, that improve the quality of the results and reduce the tensions.

\section{Future calculations}
\begin{figure}[h!]  
\centering          
\includegraphics[width=0.49\textwidth,angle=0]{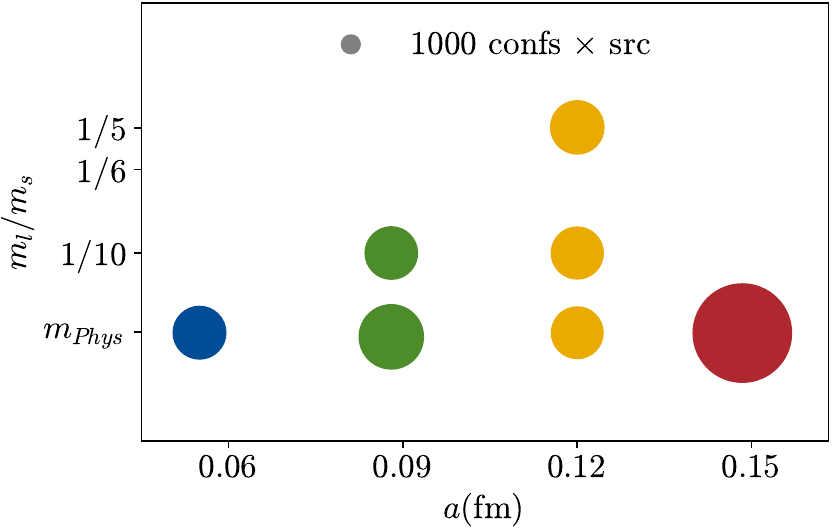}
\includegraphics[width=0.49\textwidth,angle=0]{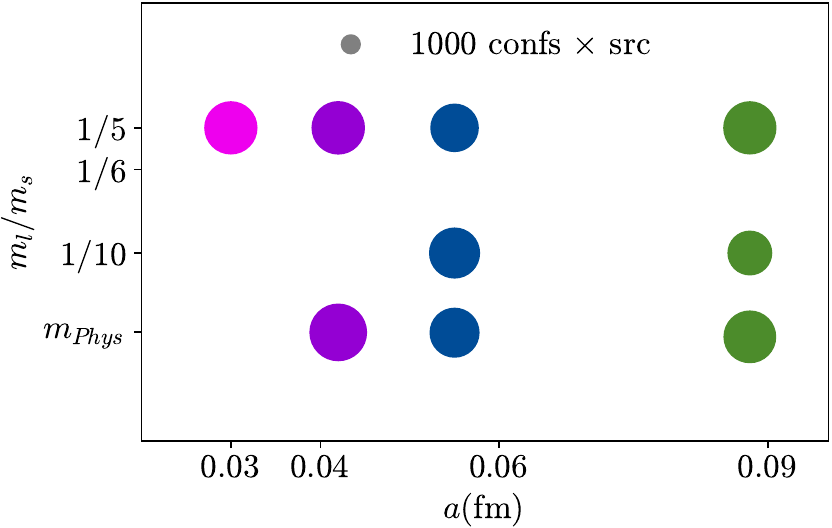}
\caption{Ensemble lists for both calculations. {\bf Left:} Calculation with Fermilab heavy quarks. {\bf Right:} Calculation with HISQ heavy quarks.}
\label{Ensembles}   
\end{figure}

The Fermilab Lattice and MILC collaborations proposed two different calculations to address current issues in the flavor physics panorama.
These two calculations cover four heavy-to-heavy channels, $B_{(s)}\to D^{(\ast)}_{(s)}\ell\nu$, and three heavy-to-light ones, namely $B\to\pi\ell\nu$, $B_s\to K\ell\nu$ and $B\to K\ell\ell$.
The calculations use $N_f=2+1+1$ HISQ ensembles, a well-known regularization due to its good properties when dealing with heavy quarks~\cite{Follana:2006rc}.
The distribution of ensembles is shown in fig.~\ref{Ensembles}.
The first calculation, on the left pane of fig.~\ref{Ensembles} uses the Fermilab interpretation~\cite{El-Khadra:1996wdx} of the clover action for the heavy quarks in the valence sector.
Since the lattice spacings for this calculation range from $0.15$ fm down to $0.06$ fm, we really need to resort to EFTs to obtain physical results.
The main improvements of this calculation with respect to the previous one~\cite{FermilabLattice:2021cdg} is the reduction of systematics associated to the chiral-continuum extrapolation, due to the usage of a better light quark action, and to the addition of several ensembles with physical pion masses.
The current timeframe of this calculation is to publish results in the following months.

The second calculation uses finer ensembles, ranging from $0.09$ fm to $0.03$ fm.
This feature, in combination with the good properties of the HISQ action, allows us to reach physical bottom quark masses without using EFTs.
We should expect then a sizable reduction of the heavy-quark discretization errors, which is the main source of systematic errors in these kind of calculations.
The timeframe of this calculation varies with the channel: the computation of the $B_{(s)}\to D_{(s)}\ell\nu$ form factors is almost complete, whereas other channels have just started or do not even have the whole dataset of correlators generated.
We should expect complete results of the seven channels in a few years.

\section{Conclusions}
The most recent heavy-to-heavy LQCD calculations have shed some light into the nature of the current $B$~anomalies in flavor physics, but we are still far from obtaining a final answer to the questions of $|V_{cb}|$ and the LFU ratios.
On the other hand, the heavy-to-light form factor calculations show discrepancies between lattice-QCD calculations from different collaborations.
Experiments are generating results at quick pace, and a failure to match those efforts from the theory side would severely limit the impact of experimental results.
                      
The lattice community is determined to deliver better results to understand these key problems in flavor physics.
In particular, the Fermilab lattice and MILC collaborations have a well-defined roadmap to address these issues, with two different calculations that will deliver high-quality results during the following months and years.

\section*{Acknowledgments}
Computations for this work were carried out with resources provided by the USQCD Collaboration; by the ALCF and NERSC, which are funded by the U.S. Department of Energy.
This work was partly supported by the Agencia Estatal de Investigación (Spain) under grant No. PID2024-160228NB-I00 (Proyectos de Generación de Conocimiento 2024).

\section*{References}
\bibliography{Moriond-PoS}
\end{document}